\documentclass[11pt]{article}
\usepackage{appb,epsfig}
\newcommand{\beq}{\begin{eqnarray}}
\newcommand{\eeq}{\end{eqnarray}}
\newcommand{\beqlet}{\begin{eqlettarray}}
\newcommand{\eeqlet}{\end{eqlettarray}}
\newcommand{\ba}{\begin{array}}
\newcommand{\ea}{\end{array}}

\def\plb#1#2#3{{\it Phys. Lett. }{\bf B#1~}(19#2)~#3}

\def\npb#1#2#3{{\it Nucl. Phys. }{\bf B#1~}(19#2)~#3}

\def\hepph#1{{\bf hep-ph}/#1}
\newcommand{\lsim}{\raisebox{-0.13cm}{~\shortstack{$<$ \\[-0.07cm] $\sim$}}~}

\def\pplus{{\mathbf{\hat p}_{\mathbf{+}}}}

\def\kplus{{\mathbf{\hat k}_{\mathbf{+}}}}

\def\soner{{\mathbf{s}^{\mathbf{*}}_{\mathbf{1}}}}
\def\stwor{{\mathbf{s}^{\mathbf{*}}_{\mathbf{2}}}}
\def\splus{{\mathbf{s}_{\mathbf{+}}}}
\def\sminus{{\mathbf{s}_{\mathbf{-}}}}

\def\pplusn{{\mathbf{p}_{\mathbf{+}}}}
\def\pminusn{{\mathbf{p}_{\mathbf{-}}}}

\def\kplusn{{\mathbf{k}_{\mathbf{+}}}}
\def\kminusn{{\mathbf{k}_{\mathbf{-}}}}

\begin{document}

\begin{flushright}
KA--TP--9--1998\\
{\tt hep-ph/9807205}\\
June 1997 \\
\end{flushright}

\def\thefootnote{\fnsymbol{footnote}}

\centerline{\large \bf 
Dipole Moments in Supersymmetry\footnote{Presented at the Zeuthen Workshop
on Elementary Particle Physics ``Loops and Legs in Gauge Theories", Rheinsberg,
April 19--24, 1998.}
}

\vspace{0.3cm}
\centerline{\sc Jos\'e I. Illana} 

\vspace{0.3cm}
\centerline{Institut f\"ur Theoretische Physik, Universit\"at Karlsruhe,}

\centerline{D--76128 Karlsruhe, Germany}

\begin{abstract}
The one--loop MSSM contributions to the weak dipole moments of the $\tau$
lepton and the $b$ quark (at the $Z$ peak) as well as the electromagnetic 
and weak dipole form factors of the $t$ quark (at arbitrary $s>4m^2_t$) are 
reviewed. Emphasis is given to the relevance of the $t$--quark 
CP--violating weak and electric dipole form factors as a part of the 
full one--loop correction to the process $e^+e^-\to t\bar t$ in the MSSM.
\end{abstract}
\PACS{12.60. Jv, 13,40. Em, 14.60. Fg, 14.65. Fy}
  
\def\thefootnote{\arabic{footnote}}
\setcounter{footnote}{0}

\section{Dipole moments}

The most general Lorentz structure of the vertex function describing the 
interaction of a neutral vector boson with on--shell fermions is
\beq
\Gamma^{Vff}_\mu={\rm i}\left\{\gamma_\mu\left[F^V_{\rm V}-F^V_{\rm A}\gamma_5
\right]+\left[{\rm i}F^V_{\rm S}+F^V_{\rm P}\gamma_5\right]p_\mu
+\left[{\rm i}F^V_{\rm M}+F^V_{\rm E}\gamma_5\right]\sigma_{\mu\nu}p^\nu
\right\}
\label{vertex}
\eeq
with $p$ being the incoming momentum of the vector boson. The
form factors $F_i$ account for the dynamics of the interaction and are 
functions of the invariant $s=p^2$. The vector and axial--vector form 
factors are the only ones that can receive a contribution
at tree level in a renormalizable theory while the others are given by
quantum corrections. 
The effects of the scalar and pseudoscalar form factors, $F^V_{\rm S}$ and 
$F^V_{\rm P}$, vanish for on--shell vector bosons and are negligible for 
the physical processes of interest (typically proportional to the electron
mass). 

The magnetic and electric form factors, $F^V_{\rm M}$ and $F^V_{\rm E}$, 
are related to the usual anomalous (weak) magnetic
dipole form factor [A(W)MDFF] and (weak) electric dipole form factors [(W)EDFF]
for $V=\gamma,Z$, respectively, through
\beq
a^V_f(s)=\frac{2m_f}{e}F^V_{\rm M}(s) & \mbox{and} &
d^V_f(s)=-F^V_{\rm E}(s).
\eeq
The electromagnetic properties of a fermion, namely the charge and the magnetic 
and electric dipole moments, are given by the static (classical) limit ($s=0$). 
In our convention for the covariant derivative they are 
\beqlet
\mbox{charge}&=&-F^\gamma_{\rm V}(0)=eQ_f, \\ 
\mbox{MDM}&=&\frac{F^\gamma_{\rm V}(0)}{2m_f}+F^\gamma_{\rm M}(0)
             =\frac{e}{2m_f}\left[-Q_f+a^\gamma_f(0)\right], \\
\mbox{EDM}&=&d^\gamma_f(0).
\eeqlet
The anomalous magnetic dipole moment is AMDM$\equiv a^\gamma_f(0)
\equiv-Q_f(g_f-2)/2$ and $g_f$ is the gyromagnetic ratio.\footnote{A massless 
neutral fermion might have a magnetic moment, by quantum corrections (3b). 
Massless neutrinos have zero magnetic moment in the SM to all orders but this 
is not necessarily true in a different theory.} 

Similarly, the weak dipole moments are defined from the $Zff$ vertex at 
$s=M^2_Z$. They do not have a classical analogy. One refers to them as 
\beqlet
\mbox{AWMDM}&=&a^w_f\equiv a^Z_f(M^2_Z), \\ 
\mbox{WEDM}&=&d^w_f\equiv d^Z_f(M^2_Z).
\eeqlet 
At the $Z$ resonance they are gauge invariant and contribute to physical 
processes with (almost) no interference with the electromagnetic form factors.

The (W)EDMs are CP violating what makes their study of particular interest.
Unlike the vector and axial--vector form factors, the dipole form factors 
are chirality flipping and therefore they must be proportional to a fermion 
mass (either in the loop or in the external legs). The heavier fermions are 
hence the best candidates to have larger dipoles. 

Here we review the MSSM contributions to the weak dipole moments of the
$\tau$ lepton and the $b$ quark (the heaviest fermions to which an on--shell 
$Z$ boson can decay) of relevance for LEP. The $t$ quark dipole form factors, 
both weak and electromagnetic, are of interest for the NLC but at $s>M^2_Z$ 
the physical observables depend on more form factors than the ones
given in the vertices $\gamma ff$ and $Zff$. For consistency, we briefly 
present the MSSM predictions for the CP violating $t$ (W)EDFFs and compare 
their influence on some CP--odd observables in the context of the full 
calculation of the process $e^+e^-\to t\bar{t}$ to one loop.

\section{One--loop weak dipole moments of \boldmath{$\tau$} and \boldmath{$b$}
 	 in the MSSM}

The $\gamma ff$ and $Zff$ one--loop vertex corrections can be classified
into six classes of topologies. Expressions in terms of generic couplings
and three--point integrals for the contribution of a renormalizable theory 
to the dipole form factors in the 't Hooft--Feynman gauge
can be found in Refs.~\cite{hirs1,hirs2}. The integrals involved are all IR 
and UV finite.
The contributions to the CP--violating (W)EDFFs are proportional to the
imaginary part of combinations of couplings.

\subsection{The magnetic dipole moments}
 
The one--loop SM prediction for the AWMDM of the $\tau$ lepton and the $t$ quark
were first calculated in \cite{ber}. The only free parameter, the SM Higgs
boson mass, does not significantly affect the electroweak contribution to 
$a^w_\tau=(2.10+0.61\ i)\times10^{-6}$ but it is more important for
the real part of $a^w_b=[(1.1;2.0;2.4)-0.2\ i]\times10^{-6}$, with
$M_{H^0}=M_Z,$ $2M_Z,$ $3M_Z$  respectively. The QCD contribution
(a gluon exchange diagram) in the case of the $b$ quark
dramatically enlarges the result to $a^w_b=(-2.96+1.56\ i)\times10^{-4}$.

In the MSSM, the different Higgs sector consists of a constrained 2HDM 
mainly controlled by the pseudoscalar Higgs boson mass $M_A$, the $\mu$ 
parameter and the ratio of VEVs $\tan\beta$. Besides, several  
soft--SUSY--breaking terms must be introduced. A simplified set is given by the 
assumption of R--parity conservation, universal scalar mass terms 
$m_{\tilde{q}}$ for squarks and $m_{\tilde{l}}$ for sleptons, trilinear terms
$A_\tau$, $A_b$ and $A_t$ (for the third family) and gaugino mass terms related
by the GUT constraint: $\alpha M_3=\alpha_s s^2_W M_2=3/5\ \alpha_s c^2_W M_1$.
In Ref.~\cite{hirs1} a complete scan of the SUSY parameter space was performed
with the following results:

\begin{itemize}

\item
The Higgs sector can provide the only contribution to the imaginary part,
of the order of the SM contribution, assuming the present experimental limits 
on the masses of the superpartners. The real part is typically negative and 
not very large.

\item
The neutralino contribution to the real part is also small and has opposite
sign than $\mu$ in most of the parameter space.

\item
The chargino contribution is the dominant one, being real and with the
same sign as $\mu$: $|$Re$(a^w_\tau)|\lsim0.2(7)\times
10^{-6}$ and $|$Re$(a^w_b)|\lsim 1(30)\times10^{-6}$ for $\tan\beta=1.6 (50)$ 
respectively.

\item
The gluinos compete in importance with the charginos for the $b$: 
$|$Re$(a^w_b)|\lsim2(40)\times10^{-6}$.

\end{itemize}

The sum of the MSSM contributions can amount to $|$Re$(a^w_\tau)|\lsim 0.5(7)
\times10^{-6}$  and $|$Re$(a^w_b)|\lsim2(50)\times 10^{-6}$
for not so extreme and not excluded regions of the SUSY parameter space. 
Decoupling is observed for large values of the parameters.

\subsection{The electric dipole moments}

In the SM there is only one source of CP violation, the $\delta_{\rm CKM}$ 
phase of the Cabibbo-Kobayashi-Maskawa (CKM) mixing matrix for quarks
(appart from $\theta_{\rm QCD}$ that will be ignored).
The SM contribution to the (W)EDM comes at three loops and hence it is as
small as $\sim e G_F m_f \alpha^2\alpha_s J/(4\pi)^5$, with $J\equiv
c_1c_2c_3s^2_1s_2s_3s_\delta$ (an invariant under reparametrizations of the
CKM matrix). That means (W)EDM$\sim3\times10^{-34}\ (10^{-33})\ e$cm for  
$\tau$ ($b$), respectively

In the MSSM there appear new physical phases, provided by the soft--breaking 
terms, whose effects show up already at the one-loop level. We restrict 
ourselves to R--parity preserving and generation--diagonal trilinear 
soft-breaking terms and assume the unification of the 
soft--breaking gaugino masses at the GUT scale. 
In such a constrained framework the following SUSY parameters can be complex:
the $\mu$ parameter, the gaugino masses, the bilinear mixing mass parameter 
$m^2_{12}$ and the trilinear soft-SUSY-breaking parameters.
Not all of these phases are physical: the MSSM has two additional U(1) 
symmetries for vanishing $\mu$ and soft--breaking terms, the Peccei--Quinn and 
the R--symmetry, that can be used to absorb two of the phases by redefinition 
of the fields \cite{relax}. In addition, the GUT constraint leads to only one 
common phase for the gaugino mass terms. Our choice of CP violating physical 
phases is:
$\varphi_\mu\equiv{\rm arg}(\mu)$, 
$\varphi_{\tilde{f}}\equiv{\rm arg}(m^f_{\rm LR})$ ($f=\tau,\ t,\ b$)
with $m^t_{\rm LR}\equiv A_t-\mu^*\cot\beta$ and 
$m^{\tau,b}_{\rm LR}\equiv A_{\tau,b}-\mu^*\tan\beta$. The scan of the SUSY
parameter space \cite{hirs2} leads to numerical results of the same order 
as the AWMDM. As a summary:

\begin{itemize}

\item
The MSSM Higgs sector is CP conserving and therefore it does not contribute to
the (W)EDM.

\item
The diagrams with neutralinos involve $\varphi_\mu$ and $\varphi_{\tilde{\tau}}$
($\varphi_{\tilde{b}}$) for the $\tau$ ($b$) case and its contribution is
maximal for these phases being $\pi/2$.

\item
The chargino diagrams only involve $\varphi_\mu$ in the $\tau$ case (there is
no scalar neutrino mixing) and also $\varphi_{\tilde{t}}$ for the $b$.
The former contribution is enhanced for $\varphi_\mu=\pi/2$ and the two
phases conspire in the latter to yield a maximum effect for $\varphi_\mu=\pi/2$
and $\varphi_{\tilde{t}}=\pi$.

\item
The gluinos contribute maximally to the $b$ WEDM for
$\varphi_{\tilde{b}}=\pi/2$.

\end{itemize}

The total one--loop predictions of the MSSM are 
$|$Re$(d^w_\tau)|\lsim 0.3(12)\times10^{-21}\ e$cm, 
$|$Re$(d^w_b)|\lsim 1.4(35)\times10^{-21}\ e$cm.

\section{One--loop dipole form factors of the \boldmath{$t$} quark
         in the MSSM}

The $t$ quark weak dipole moments ($s=M^2_Z$) cannot be defined and it is not
conceivable to measure the static electromagnetic properties ($s=0$) of such a
short--living particle. Only the electromagnetic and weak form factors
at $s>4m^2_t$ can contribute to actual processes involving top quarks. 

Since no gauge bosons (neither ghosts or Goldstone bosons) are involved in the
one--loop contribution to the (W)EDFFs, they are gauge
independent in the MSSM even for off--shell photon and $Z$ bosons. The same
is not true for the A(W)MDFFs.
 
\begin{figure}
\begin{center}
\begin{tabular}{cc}
\hspace{2.5cm}
Re$[d^\gamma_t(s)]/\mu_t$\hspace{-5cm}
\epsfig{file=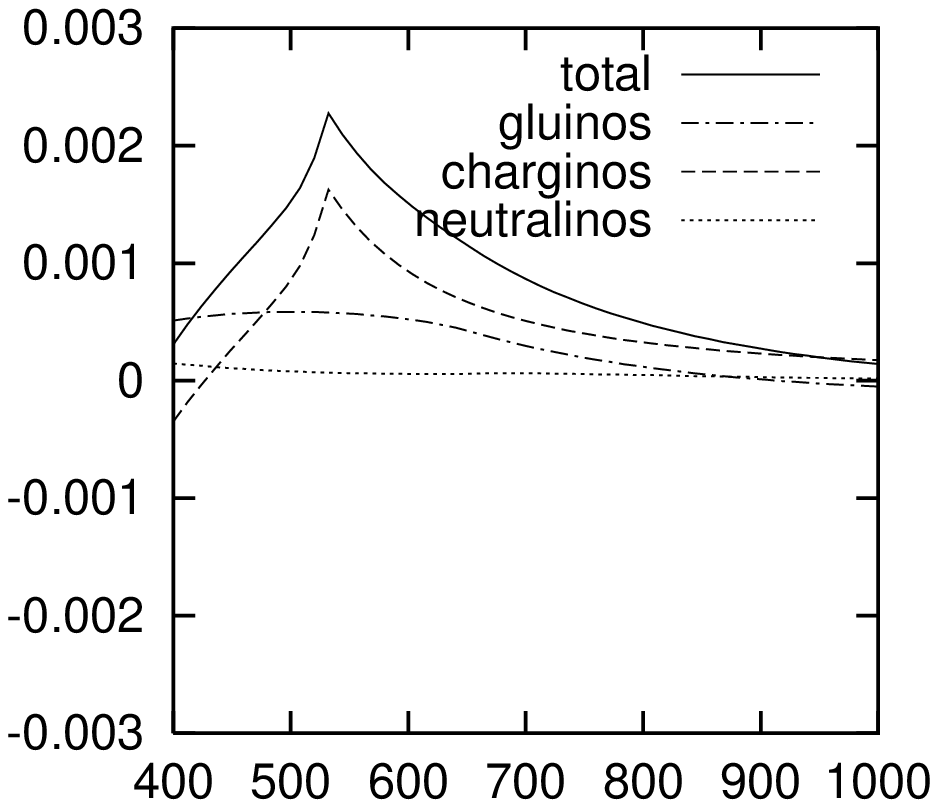,height=7cm,angle=-90} & \hspace{0.75cm}
Re$[d^Z_t(s)]/\mu_t$\hspace{-4.75cm}\epsfig{file=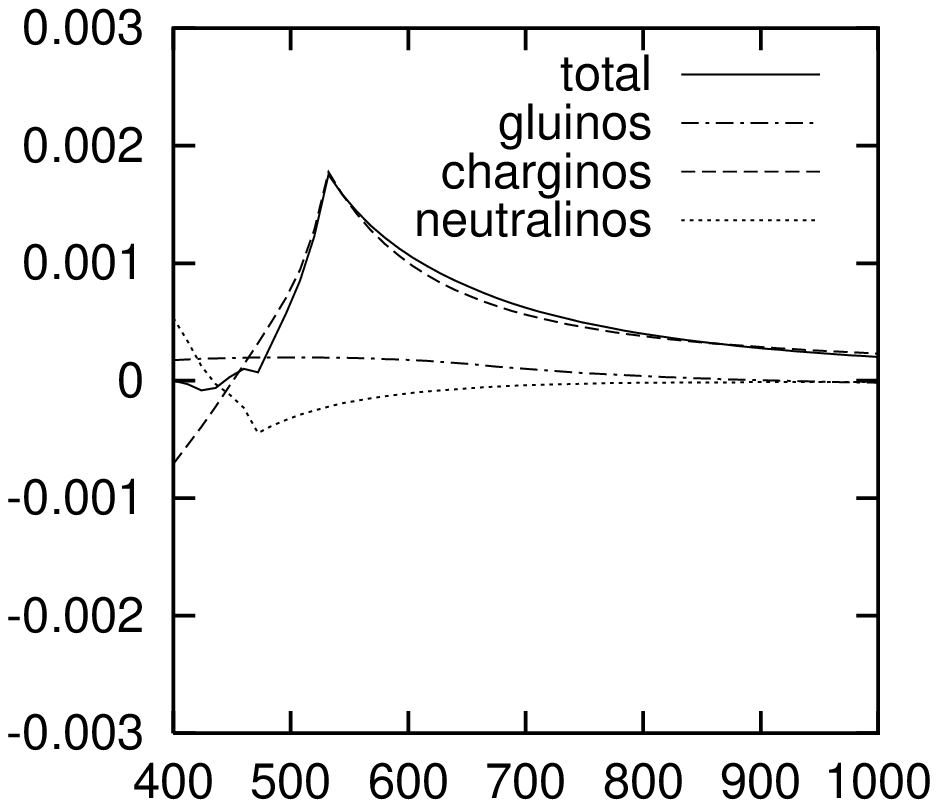,height=7cm,angle=-90} \\
\hspace{2.5cm}
Im$[d^\gamma_t(s)]/\mu_t$\hspace{-5cm}
\epsfig{file=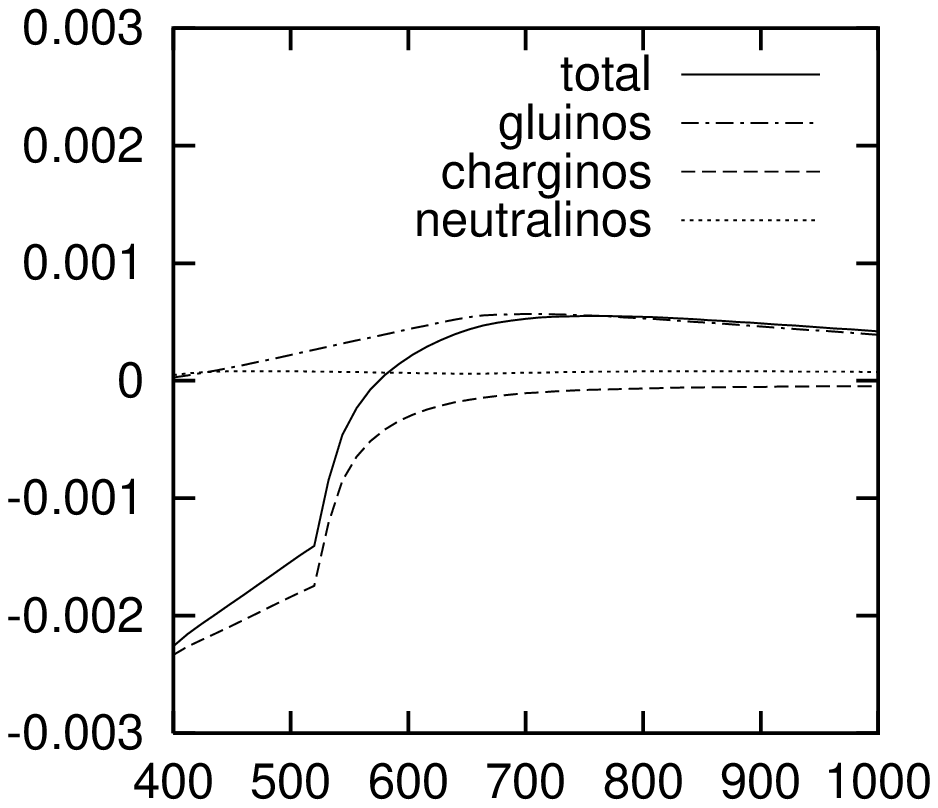,height=7cm,angle=-90} & \hspace{0.75cm}
Im$[d^Z_t(s)]/\mu_t$\hspace{-4.75cm}\epsfig{file=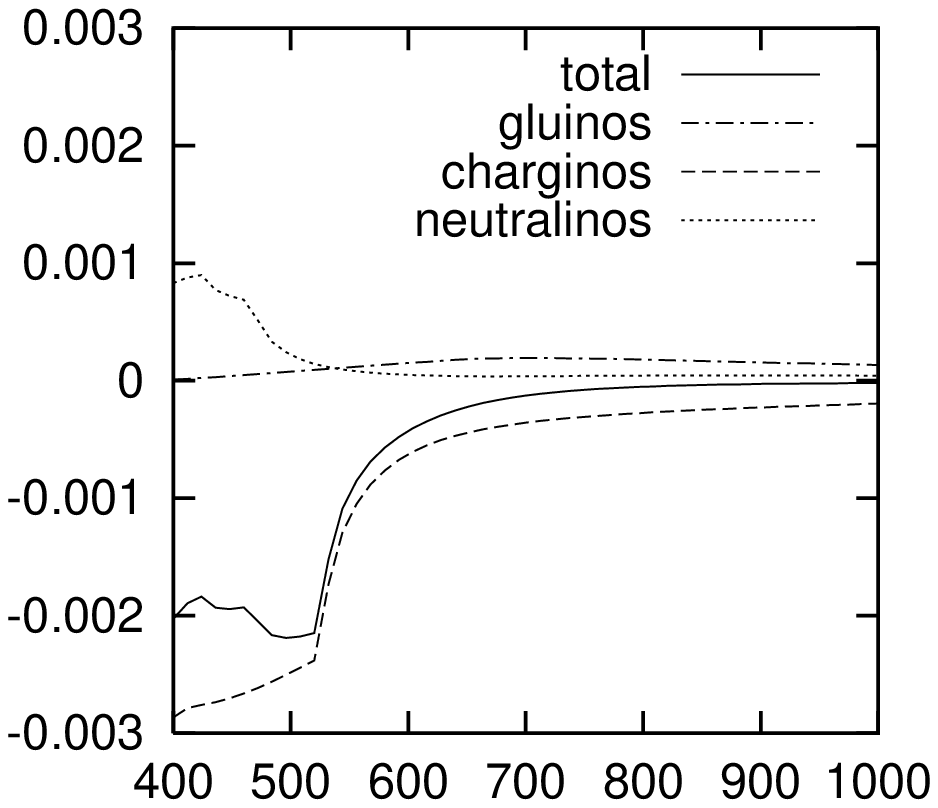,height=7cm,angle=-90} \\
  & {\small $\sqrt{s}$} 
\end{tabular}
\end{center}
\caption{The different contributions to the top (W)EDFFs for the reference set 
of SUSY parameters (\ref{susyset}) [$\mu_t=5.64\times10^{-17}\ e$cm].
\label{fig1}}
\end{figure}

The MSSM predictions for the electric and weak electric dipole 
form factors of the $t$ quark have been evaluated independently 
in Ref.~\cite{vienna}.
After a full scan of the SUSY space, a set of parameters can be found 
\cite{hirss} that maximizes the value of the $t$ (W)EDFFs. The choice of the 
optimal set closely depends on the value of the top--pair invariant mass, due 
to the possibility of threshold effects from supersymmetric
particles running in the loop. For instance, the reference set
of SUSY parameters given by
\beqlet 
\tan\beta=1.6,\ M_2=|\mu|=m_{\tilde{q}}=
|m^t_{LR}|=|m^b_{LR}|=200\ \mbox{GeV} \\ 
\mbox{and the CP--phases}\ 
\varphi_\mu=-\varphi_{\tilde{t}}=-\varphi_{\tilde{b}}=\pi/2
\eqalabel{susyset}
\eeqlet
leads to high value of the (W)EDFFs in the
neighbourhood of $\sqrt{s}=500$ GeV (Fig.~\ref{fig1}). Such a set of parameters
yields the following masses (in GeV) for the supersymmetric particles: 
$m_{\tilde{b}_1}=201.36$, $m_{\tilde{b}_2}=213.18$, 
$m_{\tilde{t}_1}=186.31$, $m_{\tilde{t}_2}=323.60$,
$m_{\tilde{\chi}^+_1}=153.47$, $m_{\tilde{\chi}^+_2}=263.36$,
$m_{\tilde{\chi}^0_1}=91.79$, $m_{\tilde{\chi}^0_2}=158.93$,
$m_{\tilde{\chi}^0_3}=202.34$, $m_{\tilde{\chi}^0_4}=266.83$,
$m_{\tilde{g}}=753.22$. Up to now, none of them have been ruled out by 
experimental searches. As the MSSM Higgs sector is irrelevant for CP--violation
the Higgs boson masses have no impact on the result.

We assume $\sqrt{s}=500$ GeV in the following, for which
\beqlet
d^\gamma_t(s)		&=&(1.515, -1.481)\times10^{-3} \ \mu_t, \\
d^Z_t(s)	   	&=&(0.682, -2.171)\times10^{-3} \ \mu_t
\label{dip}
\eeqlet
in $t$ magnetons  $\mu_t\equiv e/2m_t$ = $8.65\times 10^{-4}$ GeV$^{-1}$
= $5.64\times 10^{-17}\ e$cm.

\section{CP violation in \boldmath{$e^+e^-\to t\bar{t}$}}

Consider the pair--production of polarized tops 
$e^+(\pplusn)+e^-(\pminusn)\to t(\kplusn,\splus)+
\bar{t} (\kminusn,\sminus)$. A list of CP--odd spin observables 
classified according to their CPT properties is given in Table~\ref{tab1}.
The directions of polarization of top/antitop {\bf a}, {\bf b} are taken normal
(N), transversal (T) to the scattering plane or 
longitudinal (L). They can be either paralell ($\uparrow$) or antiparalell 
($\downarrow$) to the axes defined by $\hat{z}=\kplus$, $\hat{y}=\kplus\times
\pplus/|\kplus\times\pplus|$ and $\hat{x}=\hat{y}\times\hat{z}$. A well known
asymmetry is $\langle{\cal O}_5\rangle
=[N(t_L\bar t_L)-N(t_R\bar t_R)]/[N(t_L\bar t_L)+N(t_R\bar t_R)]$. 

\begin{figure}
\begin{center}
\begin{tabular}{lll}
\hspace{2cm}
${\cal O}_1$\hspace{-3.5cm}
\epsfig{file=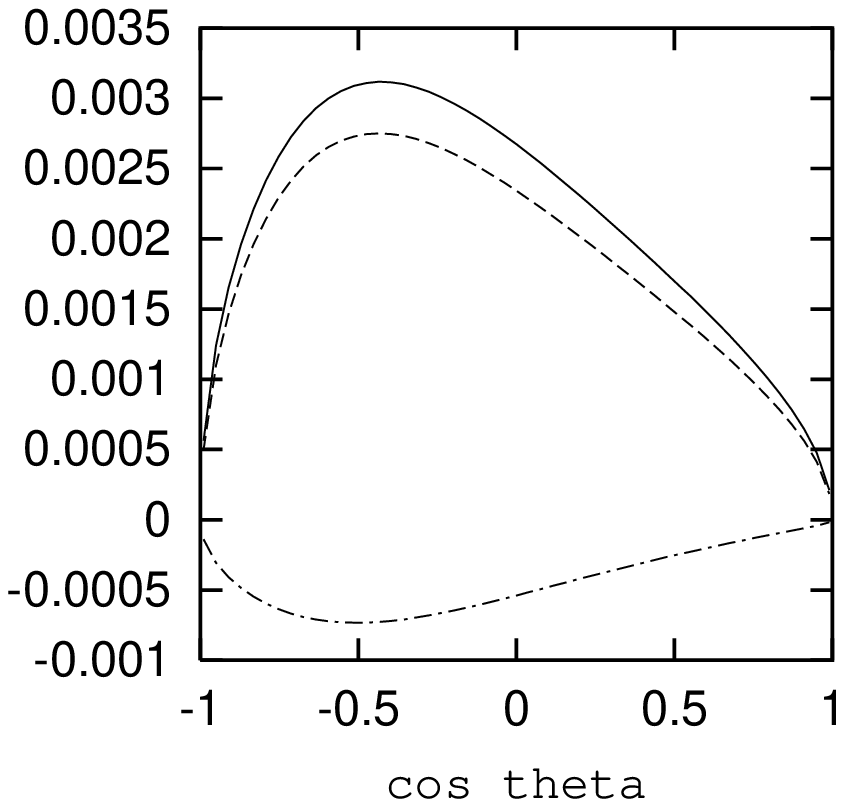,height=6cm,angle=-90} &
\hspace{0.5cm} 
${\cal O}_2$\hspace{-3.5cm}
\epsfig{file=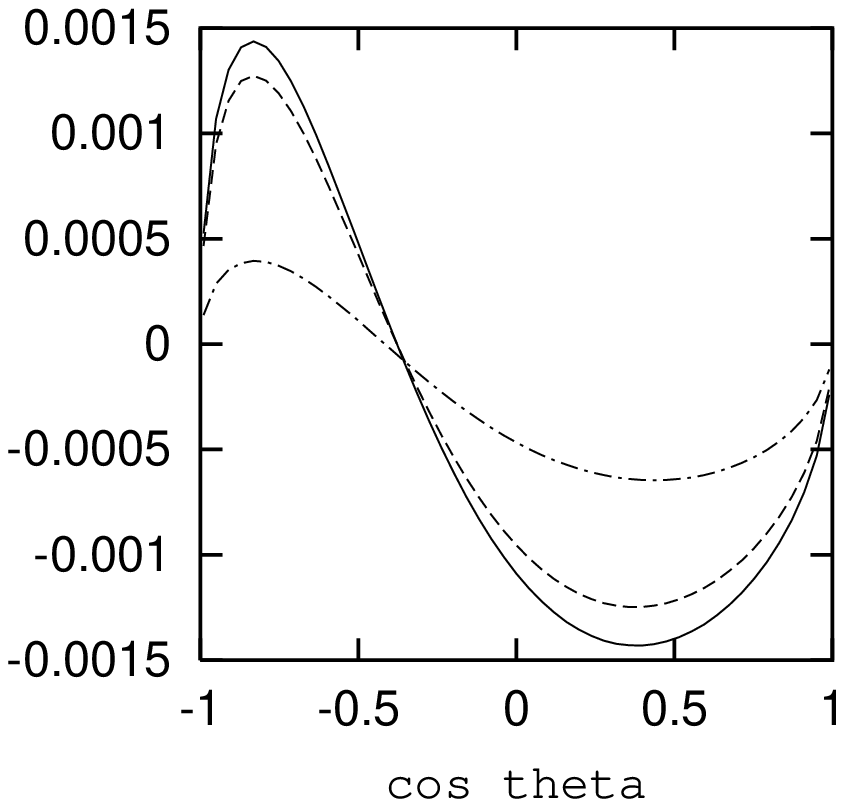,height=6cm,angle=-90} &
\hspace{0.5cm} 
${\cal O}_3$\hspace{-3.5cm}
\epsfig{file=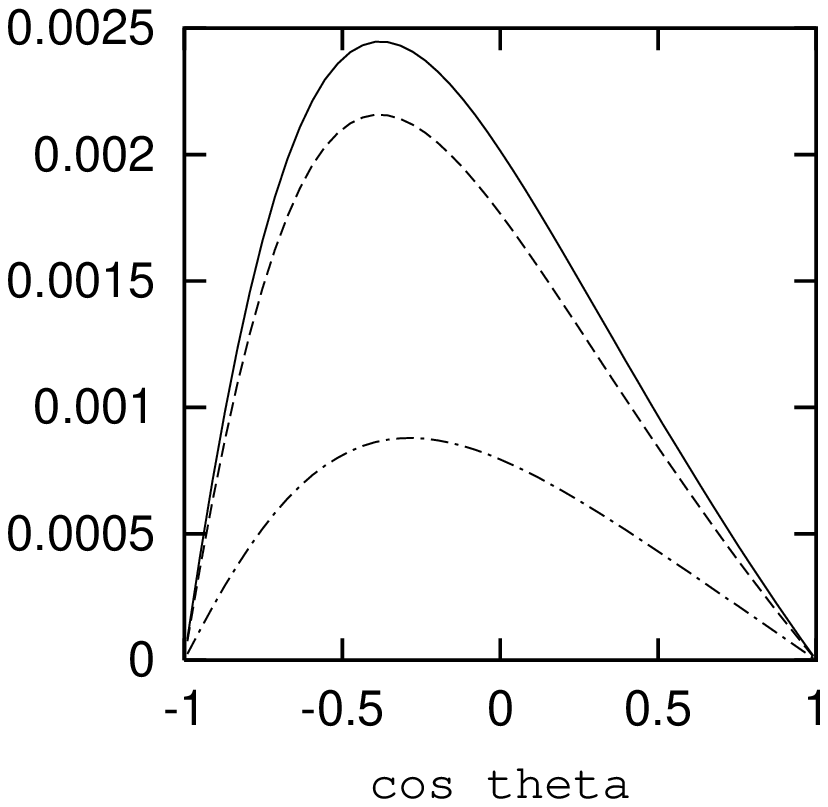,height=6cm,angle=-90} \\
\hspace{2cm}
${\cal O}_4$\hspace{-3.5cm}
\epsfig{file=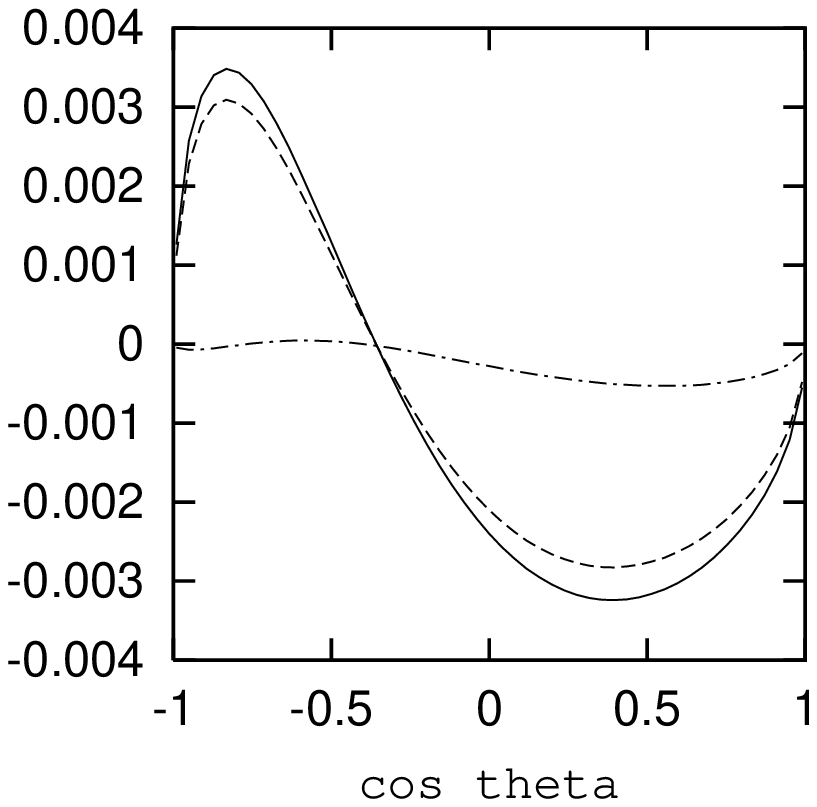,height=6cm,angle=-90} &
\hspace{0.5cm} 
${\cal O}_5$\hspace{-3.5cm}
\epsfig{file=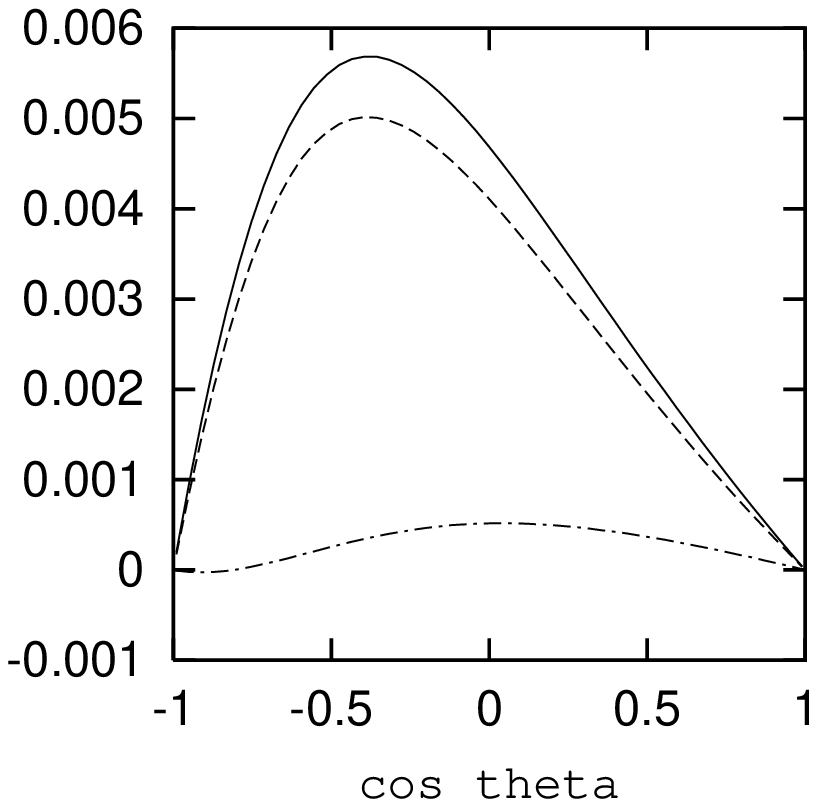,height=6cm,angle=-90} &
\hspace{0.5cm} 
${\cal O}_6$\hspace{-3.5cm}
\epsfig{file=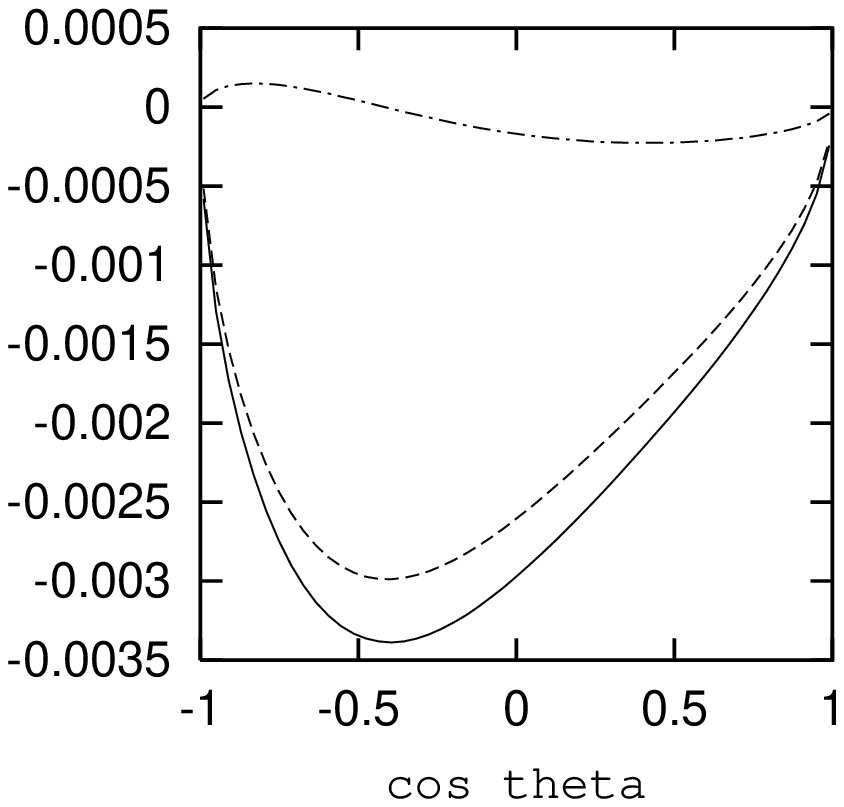,height=6cm,angle=-90}
\end{tabular}
\end{center}
\caption{The expectation value of the spin observables for the reference
 set of SUSY parameters, assuming for the cross section: 
  (i) tree level plus contribution from (W)EDFFs only (solid line);  
 (ii) one loop including all the vertex corrections and the self--energies
      (dashed line);
(iii) complete one loop (dot--dashed line).
\label{fig2}}
\end{figure}

The one--loop MSSM correction to the differential cross section for top--pair 
production \cite{hs} has been extended to accommodate complex parameters.
Two types of box graphs are involved: one with vector boson exchange
containing only (CP--even) SM contributions\footnote{
The box graphs with Higgs--boson exchange are proportional to the
electron mass and are neglected.}
($[ZeZt]$, $[W\nu_e Wb]$) and one CP--violating, purely supersymmetric 
($[\tilde e\tilde\chi^0\tilde\chi^0\tilde t]$, 
$[\tilde\nu\tilde\chi^\pm\tilde\chi^\pm\tilde b]$). 
In Fig.~\ref{fig2} we compare the contributions of dipoles and boxes to
the spin observables for the reference set of parameters (\ref{susyset})
as a function of the polar angle of the $t$ quark.
The plots show that the MSSM box graphs contribute in general to CP violation
in the process $e^+e^-\to t \bar t$ by roughly the same amount and
with a different profile than the MSSM (W)EDFFs. The sum of both
effects can generate either suppressions or enhancements of the CP signal
depending on the observables and also on the set of SUSY parameters employed.
For our choice there happens to be always a suppression.
The shape of the solid and dashed curves is the same; their different size 
is due to the contributions to the normalization factors coming from 
self--energies, A(W)MFFs and other CP--even vertex corrections.

In Table~\ref{tab1} we show the expectation values for the integrated spin 
observables and the ratio $r\equiv\langle{\cal O}\rangle/\sqrt{\langle{\cal
O}^2\rangle}$ ($|r|\sqrt{N}$ provides the statistical significance of the 
signal of CP violation for a sample of $N$ events).  
There we compare the result for (i) the assumption of tree level 
cross section including only the (W)EDFFs (left column) to (ii) the complete
one--loop calculation (right column) provided the reference set of
SUSY parameters (\ref{susyset}). The MSSM dipole and box graphs
contribute with similar importance to the CP--odd observables and with
a significance not far form the NLC capabilities (also for more realistic
observables \cite{hirss}).
  
\begin{table}
\caption{Expectation values of the integrated spin observables and the ratio
$r=\langle{\cal O}\rangle/\sqrt{\langle{\cal O}^2\rangle}$ at 
$\sqrt{s}=500$ GeV  given
the reference set of SUSY parameters (\ref{susyset}). The left column
includes only the $t$ (W)EDFF corrections  
and the right one comes from the complete one--loop cross section for 
$e^+e^-\to t\bar t$.
\label{tab1}}
\begin{center}
\begin{tabular}{|c|l|c|c|c|r|r|r|r|}
\hline
$i$ &CPT & ${\cal O}_i$ & {\bf a}&{\bf b} & 
\multicolumn{2}{|c|}{$\langle{\cal O}\rangle_{{\bf ab}}$ ($10^{-3}$)} &
\multicolumn{2}{|c|}{$r$ ($10^{-3}$)} \\
\hline \hline
1 &even & $(\soner-\stwor)_y$ & T$\uparrow$&T$\downarrow$
	& 1.832 &$-0.347$& 1.219 &$-0.230$      \\
2 &even & $(\soner\times\stwor)_x$& T$\uparrow$&L$\uparrow$	
	&$-0.747$&$-0.363$&$-0.747$&$-0.363$       \\
3 &even & $(\soner\times\stwor)_z$& N$\uparrow$&T$\uparrow$	
	& 1.171	& 0.465& 1.171	&$0.465$       \\
4 &odd  & $(\soner-\stwor)_x$     & N$\uparrow$&N$\downarrow$
	&$-1.644$&$-0.313$ &$-1.608$&$-0.309$ \\
5 &odd  & $(\soner-\stwor)_z$     & L$\uparrow$&L$\downarrow$
	& 2.722 & 0.287 & 3.263 &0.344       \\
6 &odd  & $(\soner\times\stwor)_y$& L$\uparrow$&T$\downarrow$
	&$-2.036$&$-0.126$&$-2.036$&$0.126$       \\
\hline
\end{tabular}
\end{center}
\end{table}

\samepage
\subsection*{Acknowledgements}

It is a pleasure to acknowledge W. Hollik, S. Rigolin, C. Schappacher and 
D. St\"ockinger for a very fruitful collaboration on the subject.
My especial thanks to T. Riemann and the Organizing Committee for the 
invitation to the Workshop.
This work has been supported by the Fundaci\'on Ram\'on Areces and partially 
by the Spanish CICYT under contract AEN96-1672.

\samepage

\end{document}